\documentclass[aps,prc,twocolumn,showpacs,preprintnumbers]{revtex4}

\usepackage{graphicx}
\usepackage{dcolumn}
\usepackage{bm}

\usepackage{amsmath}
\usepackage{amstext}
\usepackage{amssymb}

\begin{document}

\title{Possibility of $\Lambda\Lambda$ pairing and its dependence on
  background density in relativistic Hartree-Bogoliubov model}

\author{Tomonori Tanigawa}
\email[Email address: ]{tanigawa@tiger02.tokai.jaeri.go.jp}
\affiliation{%
  Japan Society for the Promotion of Science, Chiyoda-ku, Tokyo
  102-8471, Japan
}%
\affiliation{%
  Advanced Science Research Center, Japan Atomic Energy Research
  Institute, Tokai, Ibaraki 319-1195, Japan
}%
\altaffiliation[Corresponding address: ]{%
  Japan Atomic Energy Research Institute, Tokai, Ibaraki 319-1195,
  Japan
}%
\author{Masayuki Matsuzaki}
\email[Email address: ]{matsuza@fukuoka-edu.ac.jp}
\affiliation{%
  Department of Physics, Fukuoka University of Education, Munakata,
  Fukuoka 811-4192, Japan
}%
\author{Satoshi Chiba}
\email[Email address: ]{sachiba@popsvr.tokai.jaeri.go.jp}
\affiliation{%
  Advanced Science Research Center, Japan Atomic Energy Research
  Institute, Tokai, Ibaraki 319-1195, Japan
}%

\date{\today}

\begin{abstract}
  We calculate a $\Lambda\Lambda$ pairing gap in binary mixed matter
  of nucleons and $\Lambda$ hyperons within the relativistic
  Hartree-Bogoliubov model.  Lambda hyperons to be paired up are
  immersed in background nucleons in a normal state. The gap is
  calculated with a one-boson-exchange interaction obtained from a
  relativistic Lagrangian.  It is found that at background
  density $\rho_{N}=2.5\rho_{0}$ the $\Lambda\Lambda$ pairing gap is
  very small, and that denser background makes it rapidly
  suppressed. This result suggests a mechanism, specific to mixed
  matter dealt with relativistic models, of its dependence on the
  nucleon density. An effect of weaker $\Lambda\Lambda$ attraction on the
  gap is also examined in connection with revised information of the
  $\Lambda\Lambda$ interaction.
\end{abstract}


\maketitle

\section{Introduction}
\label{sec:intro}

Pairing correlation in hadronic matter has been attracting attention
due to close relationship between properties of neutron stars and its
interior superfluidity.
Superfluidity inside neutron stars affects, for instance, heat capacity and
neutrino emissivity. These quantities relate to the cooling processes of
neutron stars.

In neutron stars, several types of baryon pairing appear. It is
strongly believed that neutrons form the $^{1}S_{0}$ pair in the inner
crust
region~\cite{takatsuka93:_super,wambach93:_quasip,chen93:_pairin}. At
the corresponding density $10^{-3}\rho_{0} \lesssim \rho_{B} \lesssim
0.7\rho_{0}$, where $\rho_{0}$ is the saturation density of symmetric
nuclear matter, the $^{1}S_{0}$ partial wave of the nucleon-nucleon
(\textit{NN}) interaction is attractive: In infinite matter an
attraction, no matter how weak it is, brings about the BCS instability
to the ground state. This type of pairing has been most extensively
studied for decades using various models.
Also important is the $^{3}P_{2}$ neutron pairing in the outer core
region up to $\rho_{B} \sim 2\rho_{0}$. The $^{3}P_{2}$ partial wave
of the \textit{NN} interaction is attractive enough there for neutrons
to be in a superfluid
state~\cite{takatsuka93:_super,elgaroey96:_tripl}. On the other hand,
the $^{1}S_{0}$ partial wave would become repulsive there so that the
$^{1}S_{0}$ neutron pairs would disappear. Instead, the $^{1}S_{0}$
proton pairing is expected to be realized owing to its small
fraction~\cite{takatsuka93:_super,chen93:_pairin}.

In the inner core region, baryon density becomes much larger
($\rho_{B} \gtrsim 2\rho_{0}$) and various hyperons may
appear~\cite{glendenning00:_compac_stars}. Some are expected to form
pairs in the same way as the \textit{NN} pairing owing to the
attractive $^{1}S_{0}$ partial wave of the hyperon-hyperon
(\textit{YY}) interaction.
Moreover, interspecies pairing such as $\Lambda$-neutron pairing may
be realized at the total baryon densities higher than $\rho_{B} \gtrsim
4\rho_{0}$ where fractions of the two kinds of baryon are expected
to be comparable.
These kinds of pairings affect the properties of neutron stars
through, say, suppression of the hyperon direct URCA processes.
Whether $\Lambda$ hyperons are in a super state or not plays a
decisive role for the microscopic understanding of neutron stars:
$\Lambda$ hyperons in a normal state would lead to too rapid cooling
of them and force one to modify the cooling scenarios.
Conversely, one can extract information on baryonic force and inner
structure of neutron stars from these phenomena.  Studying neutron
stars thus is the driving forces for the study on baryon superfluidity.

Despite the situations, magnitude of the hyperon pairing gaps is still
uncertain. More studies are needed exploiting available information
from various sources such as the hypernuclear spectroscopy, direct
observation of neutron stars, and so on.

Our aim of this study is twofold. One is to explore an effect of Dirac
effective mass of $\Lambda$ hyperons on the $\Lambda\Lambda$ pairing
correlations in binary mixed matter composed of $\Lambda$ hyperons and
nucleons.
In this respect, recognizing the significance of covariant
representation led to the remarkable developments in nuclear/hadron
physics for the last three decades.  As is well known nowadays,
cancellation between large Lorentz scalar and vector fields provides a
proper saturation mechanism of nuclear matter on the whole. Typical
examples are the phenomenological relativistic mean field (RMF) model
and the microscopic Dirac-Brueckner-Hartree-Fock (DBHF)
approach. Especially in the latter, selfconsistent treatment of
a nucleon spinor with a bare \textit{NN} interaction brings the
saturation points predicted by the nonrelativistic BHF approach
towards the empirical one by a repulsive relativistic effect. This
selfconsistency is the key ingredient of the relativistic models.
Then, we would like to ask a following question: What does the
selfconsistency bring to superfluidity in the composite hadronic
matter?  This is an important issue on studying neutron star matter
that has complex composition of baryons using relativistic models.

The other is to investigate an impact of the recent experimental
finding on the $\Lambda\Lambda$ pairing. The KEK-PS experiment E373,
especially the ``NAGARA''
event~\cite{takahashi01:_obser_lambd_lambd_he_doubl_hyper} may explode
the ``old'' information of $\Lambda\Lambda$ interaction which has
ruled hypernuclear systems for three decades. The event unambiguously
determined the binding energy of the two $\Lambda$
hyperons $B_{\Lambda\Lambda}$ in $\substack{\phantom{\Lambda}6 \\
  \Lambda\Lambda}$He. Most importantly, it suggests that the
$\Lambda\Lambda$ interaction is weaker than it was thought before. If
this is confirmed, the new information ought to have a significant
impact on the microscopic understanding of the properties of neutron
stars. 

Unlike the \textit{NN} pairing, there are only a few studies on the
hyperon pairing. It was first studied in a nonrelativistic framework
by~\citeauthor*{balberg98:_s_lambd}~\cite{balberg98:_s_lambd}. Then
their results were applied to the study on cooling of neutron
stars~\cite{schaab98:_implic_of_hyper_pairin_for}.  They obtained the
$\Lambda\Lambda$ pairing gap in symmetric nuclear matter using an
interaction based on the \textit{G}-matrix in symmetric nuclear matter
and an approximation of nonrelativistic effective mass obtained from
single-particle energies with first order Hartree-Fock corrections,
though their motive was application to the physics of neutron
stars. Their conclusion was that the maximal pairing gap became larger
as the background density increased; at the same time, the effective
mass of $\Lambda$ hyperons became smaller. Since smaller effective mass
generally leads to smaller pairing gap, this conclusion is against
general expectations.
\citeauthor*{takatsuka99:_super_lambd_hyper_admix_neutr_star_cores}
subsequently studied the problem using two types of bare
$\Lambda\Lambda$ interactions and of hyperon core
models~\cite{takatsuka99:_super_lambd_hyper_admix_neutr_star_cores}.
Aiming at a better approximation of neutron star matter, they used the
nonrelativistic effective mass which was obtained from the
\textit{G}-matrix calculation for composite matter of neutrons and
$\Lambda$ hyperons, and was dependent on a total baryon density and a
$\Lambda$ fraction. Their gaps were somewhat smaller
than~\citeauthor{balberg98:_s_lambd}'s due to smaller effective mass
and appropriate choice of the interaction. They also showed that the
result had considerable dependence on the interactions and the hyperon
core models owing to related uncertainties.
An important thing in common to these past
studies is use of the $\Lambda\Lambda$ interactions that are too
attractive considering the consequence of the NAGARA event, which was
unavailable at that time.

Therefore, we study the $^{1}S_{0}$ $\Lambda\Lambda$ pairing
in binary mixed matter of nucleons and $\Lambda$ hyperons using 
relativistic interactions that reflects the new experimental
information for the first time. The $\Lambda$ hyperons are immersed in
pure neutron matter or symmetric nuclear matter that is treated as a
background. We use the relativistic Hartree-Bogoliubov (RHB) model in
which density-dependence of the interaction is automatically taken
into account via the Lorentz structure. The density-dependence that is
an inherent mechanism in relativistic models may lead to novel
behavior of the pairing gap:
Since pairs are formed in medium, medium effects on a
particle-particle (\mbox{p-p}) channel interaction should be
considered. In the RHB model, bare baryon masses are reduced by the
scalar mean field. This decreased mass is the Dirac effective
mass~\cite{jaminon89:_effec,glendenning00:_compac_stars}. The mass
decrease may change the pairing gap to some extent in comparison with
that obtained with the bare masses.  Although the two preceding
studies also introduced the medium effects, each had a purely
nonrelativistic origin. It has nothing to do with the Lorentz
structure and the Dirac effective mass.
We thus intend to compare with the results of the first study
by~\citeauthor{balberg98:_s_lambd} neglecting, for the time being,
complexity of $\Lambda$-$\Sigma^{0}$ mixing that probably occurs in
asymmetric nuclear matter; this mixing will be discussed in
Sec.~\ref{sec:lambda-sigma-mixing}. Besides, other constituents
predicted to exist in neutron stars and equilibration like chemical
equilibrium of neutron star matter are ignored so that we narrow down
arguments to the impact of the revision on the $\Lambda\Lambda$
pairing properties. Such a plain treatment should be taken as the very
first step of our study on the hyperon pairing with the recently
revised interactions in neutron star matter.

This paper is organized as follows: In Sec.~\ref{sec:model}, we
illustrate the Lagrangian of the system and the gap equation for the
$^{1}S_{0}$ $\Lambda\Lambda$ pairing.  In
Sec.~\ref{sec:results-discussions}, we present results of the
$\Lambda\Lambda$ pairing properties in the binary hadronic
matter. Section~\ref{sec:summ} contains a summary.

\section{Model}
\label{sec:model}

\subsection{Lagrangian}
\label{sec:lag}

Our starting model Lagrangian of the system has the following
expression:
\begin{equation}
  \label{eq:lag}
    \begin{split}
      \mathcal{L} & = \sum_{B=N,\,\Lambda}
      \bar\psi_{B}(i\gamma_\mu\partial^\mu-M_{B})\psi_{B} \\
      & {}+ \frac{1}{2}(\partial_\mu\sigma)(\partial^\mu\sigma) {}-
      \frac{1}{2}m_\sigma^2\sigma^2  {}-
      \frac{1}{4}\Omega_{\mu\nu}\Omega^{\mu\nu}
      {}+ \frac{1}{2}m_\omega^2\omega_\mu\omega^\mu \\
      & {}- \sum_{B=N,\,\Lambda} g_{\sigma
        B}\bar\psi_{B}\sigma\psi_{B}- \sum_{B=N,\,\Lambda}
      g_{\omega B}\bar\psi_{B}\gamma_\mu\omega^\mu\psi_{B} \\
      & {}+
      \frac{1}{2}(\partial_\mu\sigma^{\ast})(\partial^\mu\sigma^{\ast})
      -\frac{1}{2}m_{\sigma^{\ast}}^2{\sigma^{\ast}}^2 -
      \frac{1}{4}S_{\mu\nu}S^{\mu\nu}
      +\frac{1}{2}m_\phi^2\phi_\mu\phi^\mu \\
      & {}-g_{\sigma^{\ast}
        \Lambda}\bar\psi_{\Lambda}\sigma^{\ast}\psi_{\Lambda} -
      g_{\phi\Lambda}\bar\psi_{\Lambda}\gamma_\mu\phi^\mu\psi_{\Lambda},
    \end{split}
\end{equation}
where
$\Omega_{\mu\nu}=\partial_{\mu}\omega_{\nu}-\partial_{\nu}\omega_{\mu}$
and $S_{\mu\nu}=\partial_{\mu}\phi_{\nu}
-\partial_{\nu}\phi_{\mu}$. The symbols
$M_{N}$, $M_{\Lambda}$, $m_{\sigma}$, $m_{\omega}$,
$m_{\sigma^{\ast}}$, and $m_{\phi}$ are the mass of nucleons, $\Lambda$
hyperons, $\sigma$ bosons, $\omega$ mesons, $\sigma^{\ast}$ bosons, and
$\phi$ mesons, respectively. Table~\ref{tab:param} displays these
masses, coupling constants and their ratios used in this study.
This model Lagrangian was used in the study of binding energy of
double $\Lambda$ hypernuclei with the RMF
model~\cite{marcos98:_bindin_lambd}, and originally proposed
by~\citeauthor{schaffner93:_stran_hadron_matter} in the study of
multiply strange hadronic systems including baryon species $N$,
$\Lambda$, and
$\Xi$~\cite{schaffner93:_stran_hadron_matter,schaffner94:_multip_stran_nuclear_system}.
The parameter set was determined
by~\citeauthor{marcos98:_bindin_lambd} to reproduce the bulk
properties of hadronic matter and finite nuclear systems including
double $\Lambda$ hypernuclei according to the old information.
In Ref.~\cite{marcos98:_bindin_lambd}, the original nucleon Lagrangian
HS (an abbreviation of Horowitz-Serot) with ``$\sigma$-$\omega$''
Lagrangian for the $\Lambda$ sector, which is called ``model~1,'' was
supplemented with two additional boson fields. One is a
scalar-isoscalar boson $\sigma^{\ast}$ (975 MeV) and the other is a
vector-isoscalar meson $\phi$ (1020 MeV). The Lagrangian that contains
$\sigma^{\ast}$ and $\phi$ is referred to as ``model~2.'' Following
Ref.~\cite{schaab98:_implic_of_hyper_pairin_for}, we call the model
Lagrangian Eq.~\eqref{eq:lag} ``HS-m2'' in short from now on. Details
of the model~1 and the model~2 are described in
Ref.~\cite{schaffner94:_multip_stran_nuclear_system}.
These additional bosons were originally introduced to achieve strong
attraction between $\Lambda$ hyperons.
Since it is, however, probable now that $\Lambda\Lambda$ interaction is
weaker than it was considered as mentioned above, we regard
$\sigma^{\ast}$ as the device for controlling the $\Lambda\Lambda$
attraction in this study.
As a rough guide, we refer to Fig.~1 of
Ref.~\cite{marcos98:_bindin_lambd} that shows the dependence of the
bond energy,
\begin{equation}
  \label{eq:bond-energy}
  \Delta B_{\Lambda\Lambda} =
  B_{\Lambda\Lambda}(\substack{\phantom{\Lambda}A \\ \Lambda\Lambda}Z)
  - 2B_{\Lambda}(\substack{A-1 \\ \phantom{A-}\Lambda}Z),
\end{equation}
on the coupling ratio
$\alpha_{\sigma^{\ast}}=g_{\sigma^{\ast}\Lambda}/g_{\sigma N}$.
Contrary, the ratio $\alpha_{\phi}=g_{\phi\Lambda}/g_{\omega N}$ is
fixed by the $SU(6)$ relations.

\begin{table}[tbp]
  \centering
  \begin{ruledtabular}
  \begin{tabular}{cdcc}
    \multicolumn{2}{c}{mass [MeV]} & \multicolumn{2}{c}{coupling constant
    $g$/ratio $\alpha$} \\
    \hline
    $m_{\sigma}$ & $520.0$ & $g_{\sigma N}$ & $10.481$ \\
                 &       & $\alpha_{\sigma}=g_{\sigma
    \Lambda}/g_{\sigma N}$ & $0.623$ \\
    $m_{\omega}$ & $783.0$ & $g_{\omega N}$ & $13.814$ \\
                 &       & $\alpha_{\omega}=g_{\omega
    \Lambda}/g_{\omega N}$ & $2/3$ \\
    $m_{\sigma^{\ast}}$ & $975.0$ &
    $\alpha_{\sigma^{\ast}}=g_{\sigma^{\ast}\Lambda}/g_{\sigma N}$
    & varied \\
    $m_{\phi}$ & $1020.0$ & $\alpha_{\phi}=g_{\phi\Lambda}/g_{\omega N}$
    & $- \sqrt{2}/3$ \\
  \end{tabular}
  \end{ruledtabular}
  \caption{Parameter set HS-m2~\cite{marcos98:_bindin_lambd}. We
    choose $M_{N}=938.0$ MeV and $M_{\Lambda}=1115.6$ MeV.}
  \label{tab:param}
\end{table}

\subsection{gap equation}
\label{sec:gap-equation}

Next, we explain the gap equation for the $\Lambda\Lambda$
pairing. The equations of motion are solved by the procedure
illustrated in Ref.~\cite{schaffner96:_hyper}, except that pairing
correlation is introduced by the Gor'kov factorization and hyperons
other than $\Lambda$ are absent in the present study. The Fock
contribution is neglected and the so-called no-sea approximation is
employed. This is the RHB model. As for the pairing gap, the gap
equation,
\begin{eqnarray}
  \label{eq:gapeq}
  \Delta (p) & = & \displaystyle {}- \frac{1}{8 \pi^{2}}
  \int_{0}^{\infty}
  \frac{\Delta (k)}{\sqrt{(E_{k}^{(\Lambda)} -
      E_{k_\mathrm{F}}^{(\Lambda)})^{2} 
      + \Delta^{2} (k)}} \nonumber \\
  & & \qquad \qquad \quad \times \;
  \bar{v}(M_{\Lambda}^{\ast};\, p, k)
  \: k^{2} dk, 
\end{eqnarray}
is solved numerically, where $E_{k}^{(\Lambda)}$ is the
single-particle energy of $\Lambda$ hyperons and
$\bar{v}(M_{\Lambda}^{\ast};\, p, k)$ is the \mbox{p-p} channel
$\Lambda\Lambda$ interaction.
Although the quality of bare $\Lambda\Lambda$ interactions on the
market is steadily getting higher, they still have room for
improvement mainly due to sparsity of experimental data. Hence, also
does predictability of the $\Lambda\Lambda$ pairing
properties. Following the same prescription as in our previous studies
of \textit{NN}
pairing~\cite{tanigawa99:_const_effec_pair_wave_funct,matsuzaki01:_phenom},
we therefore use the phenomenological interaction to study the
possibility of the $\Lambda\Lambda$ pairing and its dependence on the
density of background matter. We adopt to the \mbox{p-p} channel
interaction the one-boson-exchange (OBE) interaction obtained from an
RMF parameter set with the help of form factors. For convenience, we
refer to this OBE interaction as ``RMF interaction'' in this study.
The antisymmetrized matrix element of the RMF interaction $V$ is
defined by
\begin{equation}
\label{eq:antisym-vpp}
\bar{v}(M_{\Lambda}^{\ast};\, \mathbf{p},\mathbf{k})
=\langle \mathbf{p}s',\widetilde{\mathbf{p}s'}\vert V\vert
           \mathbf{k}s,\widetilde{\mathbf{k}s}\rangle
   -\langle \mathbf{p}s',\widetilde{\mathbf{p}s'}\vert V\vert
           \widetilde{\mathbf{k}s},\mathbf{k}s\rangle ,
\end{equation}
where tildes denote time reversal. Since $\bar{v}$
depends on the Dirac effective mass of $\Lambda$ hyperons in a baryon
spinor, its dependence is explicitly indicated in
Eqs.~\eqref{eq:gapeq} and~\eqref{eq:antisym-vpp}. Integration with
respect to the angle between $\mathbf{p}$ and $\mathbf{k}$ has been
performed in Eq.~\eqref{eq:gapeq} to project out the \textit{S}-wave
component. Moreover, the form factors are included in $\bar{v}$ to
regulate its high-momentum contributions.
We use a Bonn-type form factor:
\begin{equation}
  f(\mathbf{q}^{2})=
  \frac{\Lambda_\mathrm{c}^{2}-m_{i}^{2}}
  {\Lambda_\mathrm{c}^{2}+\mathbf{q}^{2}}\, ,
  \label{eq:bonnff}
\end{equation}
where $\mathbf{q}$ is three-momentum transfer and $m_{i}$ ($i=\sigma$,
$\omega$, $\sigma^{\ast}$, and $\phi$) are the meson masses. The
cutoff mass $\Lambda_\mathrm{c}$ is $7.26$ fm$^{-1}$ for all the
mesons employed, which value was determined in our study of the
\textit{NN} pairing for a form factor of a type different from
Eq.~\eqref{eq:bonnff}~\cite{matsuzaki01:_phenom}; an effect of varying
the cutoff mass will be examined later. We thereby aim at
phenomenological construction of the effective $\Lambda\Lambda$
interaction usable in hadronic matter with a finite $\Lambda$
fraction, like the Gogny force in the nucleon sector. Note that the
form factors are applied only to the p-p channel since we respect the
fact that the Hartree part is unaffected by a monopole form factor
(Eq.~\eqref{eq:monoff} shown later) with which the value of
$\Lambda_\mathrm{c}$ was determined in our study of the \textit{NN}
pairing.

Combining the equations for the Dirac effective mass of nucleons
\begin{eqnarray}
  \label{eq:nefmeq}
      M_{N}^{\ast} & \; = \; & M_{N}
      +g_{\sigma N}\langle\sigma\rangle,
\end{eqnarray}
that of $\Lambda$ hyperons
\begin{eqnarray}
  \label{eq:lefmeq}
      M_{\Lambda}^{\ast} & \; = \; & M_{\Lambda}
      +g_{\sigma\Lambda}\langle\sigma\rangle
      +g_{\sigma^{\ast} \Lambda}\langle\sigma^{\ast}\rangle,
\end{eqnarray}
and Eq.~\eqref{eq:gapeq}, we obtain the coupled
equations to be solved numerically.

Concerning choice of the \mbox{p-p} channel interaction, it has been
still open to argument whether the \textit{G}-matrix or a bare interaction
is suitable for the gap equation. The \mbox{p-p} channel interaction
in Ref.~\cite{balberg98:_s_lambd} is classified into the former and
that in
Ref.~\cite{takatsuka99:_super_lambd_hyper_admix_neutr_star_cores} into
the latter. As described in Ref.~\cite{migdal67:_theor_fermi}, the gap
equation itself handles short range correlations which the
\textit{G}-matrix also does; this is the well-known double-counting of
the correlations.
\citeauthor*{takatsuka99:_super_lambd_hyper_admix_neutr_star_cores}
also argued that use of the \textit{G}-matrix itself or the interaction
based on it in the gap equation cannot be
justified~\cite{takatsuka99:_super_lambd_hyper_admix_neutr_star_cores}.
Thus it is widely received to use a bare interaction in the gap equation
with regard to \textit{NN} pairing in infinite matter.

Meanwhile, from a practical viewpoint such as application to finite
nuclei, the Gogny force is often used as the \mbox{p-p} channel
interaction as well as the particle-hole channel interaction: While it
is regarded as a reasonable parameterization of the \textit{G}-matrix in
the sense that it gives saturation properties of symmetric nuclear
matter, it can reproduce the pairing gaps obtained from bare
\textit{NN} interactions. In fact, the Gogny force imitates bare
interactions in the low-density
limit~\cite{bertsch91:_pair_correl_neutr_drip_line,garrido99:_effec}.
With reference to practical usage, \citeauthor*{matsuzaki01:_phenom}
solved the gap equation of \textit{NN} pairing using the RMF
interactions~\cite{tanigawa99:_const_effec_pair_wave_funct,matsuzaki01:_phenom}.
In Ref.~\cite{matsuzaki01:_phenom}, phenomenological form factors were
introduced to the \mbox{p-p} channel, with the Hartree part unchanged,
so that the constructed interactions reproduced the results obtained
from the \emph{bare} Bonn potential. They successfully adjusted cutoff
masses to reproduce both the pairing gaps and the coherence lengths at
the same time. The cutoff masses independent of density were
qualitatively similar to those of the Bonn potential. This shows the
usefulness of the prescription. Nonetheless, we do not use the same
procedure in the present study since we have no model of the
\textit{YY} interactions to follow and our primary interest is
behavior of the gap; we use the cutoff mass for the $\Lambda\Lambda$
interaction obtained from the study of the \textit{NN}
pairing~\cite{matsuzaki01:_phenom} instead, which is qualitatively
similar to those of the \textit{YY} interactions on the market.

Last but not least, we neglect effects beyond the mean-field
approximation such as the dispersive
effects~\cite{bozek99:_super,baldo00:_disper} and the polarization
effects~\cite{ainsworth89:_effec_inter_energ_gaps_low,schulze96:_medium}
to concentrate on the effects which stem from the binary character of
the matter. Both are, however, very important for the \textit{NN}
pairing since they turned out to reduce the pairing gap. Work in this
direction is necessary in future.

\section{Results and discussions}
\label{sec:results-discussions}

\subsection{Effect of Dirac effective mass decrease}
\label{sec:effect-dirac-mass}

\begin{figure}[tbp]
  \centering
  \includegraphics[bb=0 270 592
  742,width=9cm,keepaspectratio,clip]{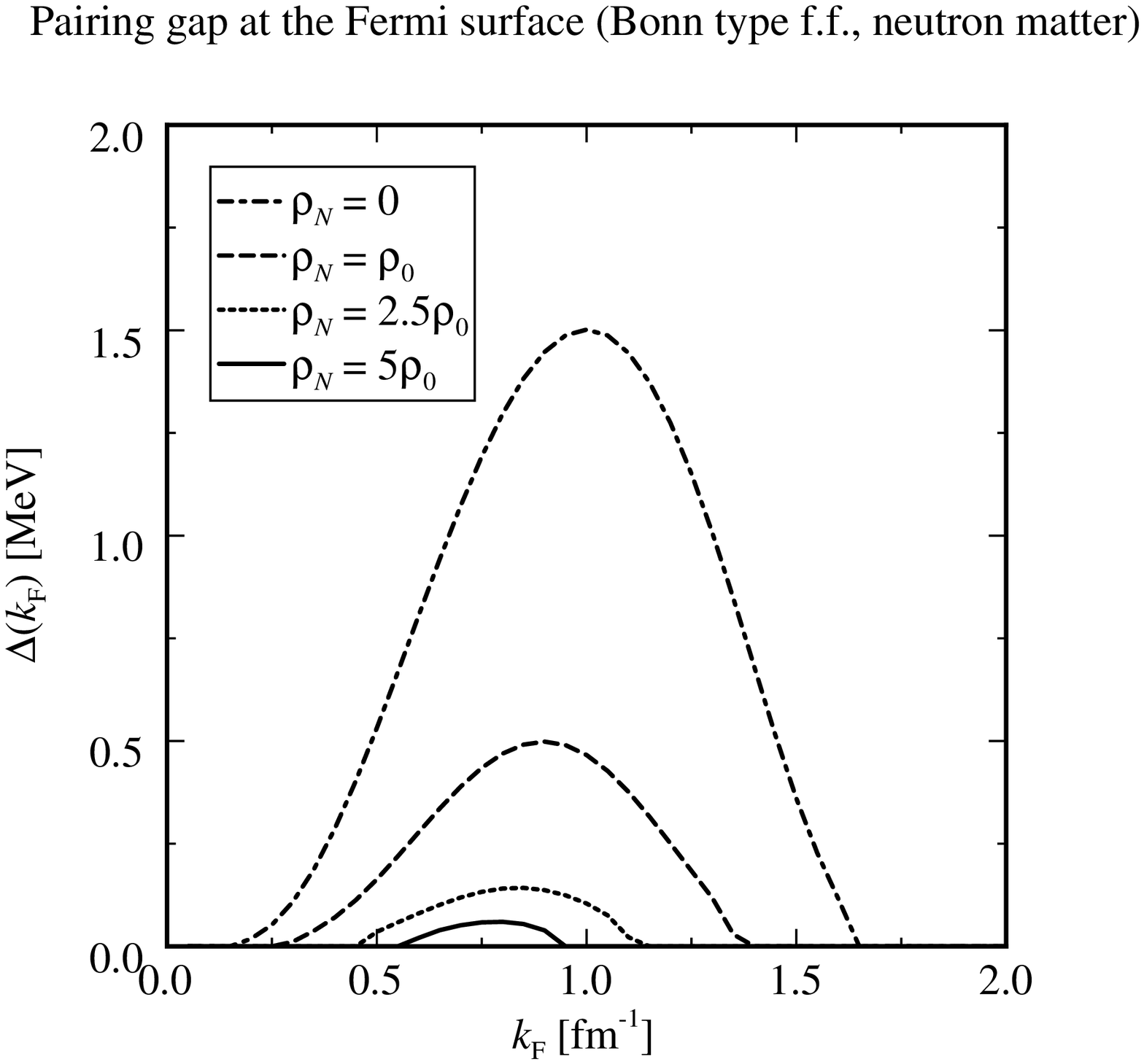} 
  \caption{$\Lambda\Lambda$ pairing gap at the Fermi surface of
  $\Lambda$ hyperons, for pure neutron background densities
  $\rho_{N}=0$, $\rho_{0}$, $2.5\rho_{0}$, and $5\rho_{0}$. The
  coupling ratio $\alpha_{\sigma^{\ast}}=0.5$ is used.}
  \label{fig:neutgap}
\end{figure}

Figure~\ref{fig:neutgap} shows the resulting $^{1}S_{0}$
$\Lambda\Lambda$ pairing gap at the Fermi surface in pure neutron
matter of densities $\rho_{N}$ at $0$, $\rho_{0}$, $2.5\rho_{0}$, and
$5\rho_{0}$, with $\alpha_{\sigma^{\ast}}=0.5$ chosen. This value of
the coupling ratio can reproduce the bond energy,
Eq.~\eqref{eq:bond-energy}, of about $1$ MeV in the RMF model, which
is suggested by the NAGARA event. Contrary to the results obtained
by~\citeauthor*{balberg98:_s_lambd}, the $\Lambda\Lambda$ pairing gap
becomes suppressed as the neutron density increases. At
$\rho_{N}=2.5\rho_{0}$, where $\Lambda$ hyperons already appear in
some models of neutron stars~\cite{schaffner96:_hyper}, the maximal
pairing gap is about $0.15$ MeV. Since there are probably no $\Lambda$
hyperons at $\rho_{N}=0$ and $\rho_{0}$ in neutron star matter, the
pairing gaps at these densities are quite hypothetical.

\begin{figure}[tbp]
  \centering
  \includegraphics[bb=0 265 592
  700,width=9cm,keepaspectratio,clip]{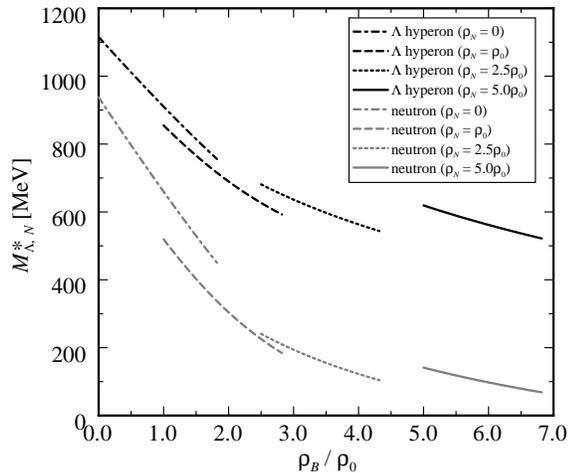}    
  \caption{Effective masses of $\Lambda$ hyperons and neutrons for pure
  neutron background densities $\rho_{N}=0$, $\rho_{0}$,
  $2.5\rho_{0}$, and $5\rho_{0}$. The coupling ratio
  $\alpha_{\sigma^{\ast}}=0.5$ is used.}
  \label{fig:neuteffmass}
\end{figure}

Figure~\ref{fig:neuteffmass} shows the density dependence of baryon
effective masses, $M_{N}^{\ast}$ and $M_{\Lambda}^{\ast}$, as
functions of the total baryon density $\rho_{B}$. The background
neutron densities are fixed here, so that variations in $\rho_{B}$
correspond to those in the Fermi momentum of $\Lambda$ hyperons. Since
we ignore the chemical equilibrium here, curves of the effective
masses have discontinuous jumps; each piece corresponds to the fixed
neutron densities, $\rho_{N}= 0$, $\rho_{0}$, $2.5\rho_{0}$, and
$5.0\rho_{0}$. Consideration of the chemical equilibrium should
connect them with each other. Nevertheless, we obtain the values
qualitatively similar to the ones shown in, for example, Fig.~4 of
Ref.~\cite{schaffner96:_hyper}. It is therefore concluded that
in-medium property of the phenomenological $\Lambda\Lambda$
interaction used in this study is justifiable. The effective mass of
neutrons decreases steeply as the total baryon density increases, while
mildly does the effective mass of $\Lambda$ hyperons due to the weaker
coupling of $\Lambda$ hyperons to the scalar bosons than the coupling
of nucleons.

\begin{figure}[tbp]
  \centering
  \includegraphics[bb=0 270 592
  742,width=9cm,keepaspectratio,clip]{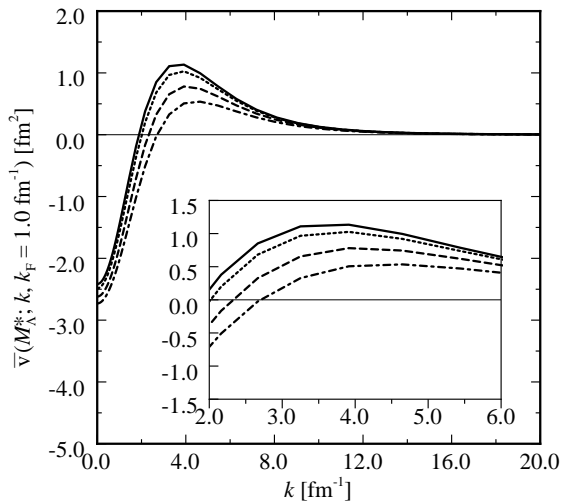}
  \caption{$\Lambda\Lambda$ RMF interaction $\bar{v}(M_{\Lambda}^{\ast};\,
    k,k_\mathrm{F})$ at the Fermi momentum of $\Lambda$
    hyperons $k_\mathrm{F}=1.0$ fm$^{-1}$, for pure neutron
    background densities $\rho_{N}=0$, $\rho_{0}$, $2.5\rho_{0}$, and
    $5\rho_{0}$, corresponding to $M_{\Lambda}^{\ast}=1068$, $813$,
    $660$, and $605$ MeV, respectively. The coupling ratio
    $\alpha_{\sigma^{\ast}}=0.5$ is used. The legend is the same as in
    Fig.~\ref{fig:neutgap}. The inset shows a magnification of the region
    around the repulsive bumps.}
  \label{fig:mexint}
\end{figure}

For the \mbox{p-p} channel, we use the RMF interaction as stated
above. The interaction contains the Dirac effective mass of $\Lambda$
hyperons, Eq.~\eqref{eq:lefmeq}, through which the medium effects are
introduced; the coupling of $\Lambda$ hyperons to $\sigma$ bosons, to
which nucleons also couple, brings about the dependence on the
background density.  Figure~\ref{fig:mexint} represents the
$\Lambda\Lambda$ RMF interaction derived from the parameter set
HS-m2. It is shown that increasing the background neutron density
suppresses attractive contribution from low momenta. This is the main
reason why the $\Lambda\Lambda$ pairing gap is smaller in denser
background.
This new mechanism of the suppression is inherent in relativistic
models which respect the Lorentz structure as shown in
Eq.~\eqref{eq:lefmeq}. It is shown that the decrease of the effective
baryon mass plays an indispensable role when it is used
selfconsistently in the baryon spinor.

What is important is that relativistic models naturally lead to a
density-dependent interaction through a selfconsistent baryon spinor
where the bare mass in a free spinor is replaced with the Dirac
effective mass. 
An apt example is the saturation of symmetric nuclear matter in the
DBHF approach~\cite{machleidt89:_meson}. Requirement of the
selfconsistency for the nucleon spinor, that is, use of the Dirac
effective mass in the nucleon spinor effectively gives repulsion to
the binding energy of symmetric nuclear matter. Consequently, it
pushes the saturation points predicted by nonrelativistic models
toward the empirical one. It seems that our finding is similar to this
repulsive effect. Furthermore, the mechanism is apparently not
restricted to $\Lambda\Lambda$ pairs. It is probable that other kinds
of \textit{YY} pairs have the same trend.

\subsection{Effect of NAGARA event}
\label{sec:effect-nagara}

\begin{figure}[tbp]
  \centering
  \includegraphics[bb=132 493 452
  675,width=8cm,keepaspectratio,clip]{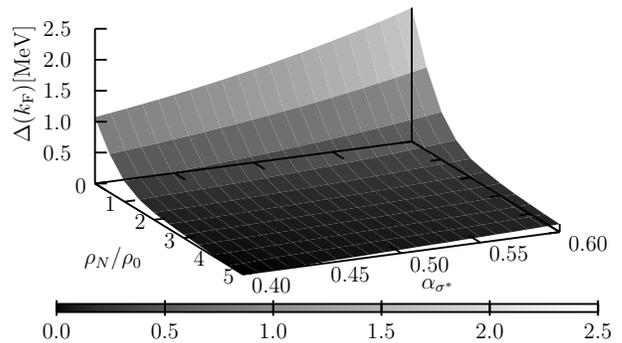}
  \caption{Maximal $\Lambda\Lambda$ pairing gap as a function of the
  strength of $\Lambda\Lambda$ attraction and the background density
  in pure neutron matter.}
  \label{fig:3dim}
\end{figure}

Next we explore the effect of the NAGARA event on the $\Lambda\Lambda$
pairing.  With relation to the revised information on the
$\Lambda\Lambda$ interaction, we vary the ratio
$\alpha_{\sigma^{\ast}}=g_{\sigma^{\ast} \Lambda} / g_{\sigma N}$
between 0.4 and 0.6, referring to Fig.~1 of
Ref.~\cite{marcos98:_bindin_lambd}: Thereby we control the attractive
component of the interaction. Figure~\ref{fig:3dim} represents the
maximal $\Lambda\Lambda$ pairing gap at the Fermi surface of $\Lambda$
hyperons as a function of the strength of $\Lambda\Lambda$ attraction
and the background density of pure neutron matter.
From this figure as well as Fig.~\ref{fig:neutgap}, one reads that the
suppression of the gap occurs in denser background of
neutrons. Moreover, it may even vanish in the end (though the result
depends on the choice of RMF parameter sets and a cutoff mass as will
be shown later).  This varying $\alpha_{\sigma^{\ast}}$ reveals likely
closing of the gap at smaller $\alpha_{\sigma^{\ast}}$ (\textit{i.e.}
weaker $\Lambda\Lambda$ attraction) and its strong suppression in the
denser neutron background. This result implies that the aforementioned
mechanism acts in concert with the weakened attraction for closing the
gap. Hence, the $\Lambda\Lambda$ pairing correlation in dense pure
neutron matter becomes less likely than before. The result is almost
the same with the background of symmetric nuclear matter.

In the light of the neutron star cooling, the absence of the
$\Lambda\Lambda$ pairing might call for the pairing of other hyperonic
species and a modification of its scenarios.
More realistic approximation of the internal composition of neutron
stars needs a condition of chemical equilibrium which plays a decisive
role. Under the condition, other hyperons will emerge as the
background density increases.
\citeauthor{takatsuka01:_possib_hyper_super_neutr_star_cores}  studied
the $\Sigma^{-}\Sigma^{-}$ and $\Xi^{-}\Xi^{-}$ pairings and shown
their
possibility~\cite{takatsuka01:_possib_hyper_super_neutr_star_cores}.
Nevertheless, the possibility of the $\Lambda\Lambda$ pairing in
neutron star matter with the concerted mechanism in this model stands
unsettled due to complex composition of particles inside neutron stars.

\subsection{Comparison with nonrelativistic study}
\label{sec:comp-with-nonr}

\begin{figure}[tbp]
  \centering
  \includegraphics[bb=0 270 592
  742,width=9cm,keepaspectratio,clip]{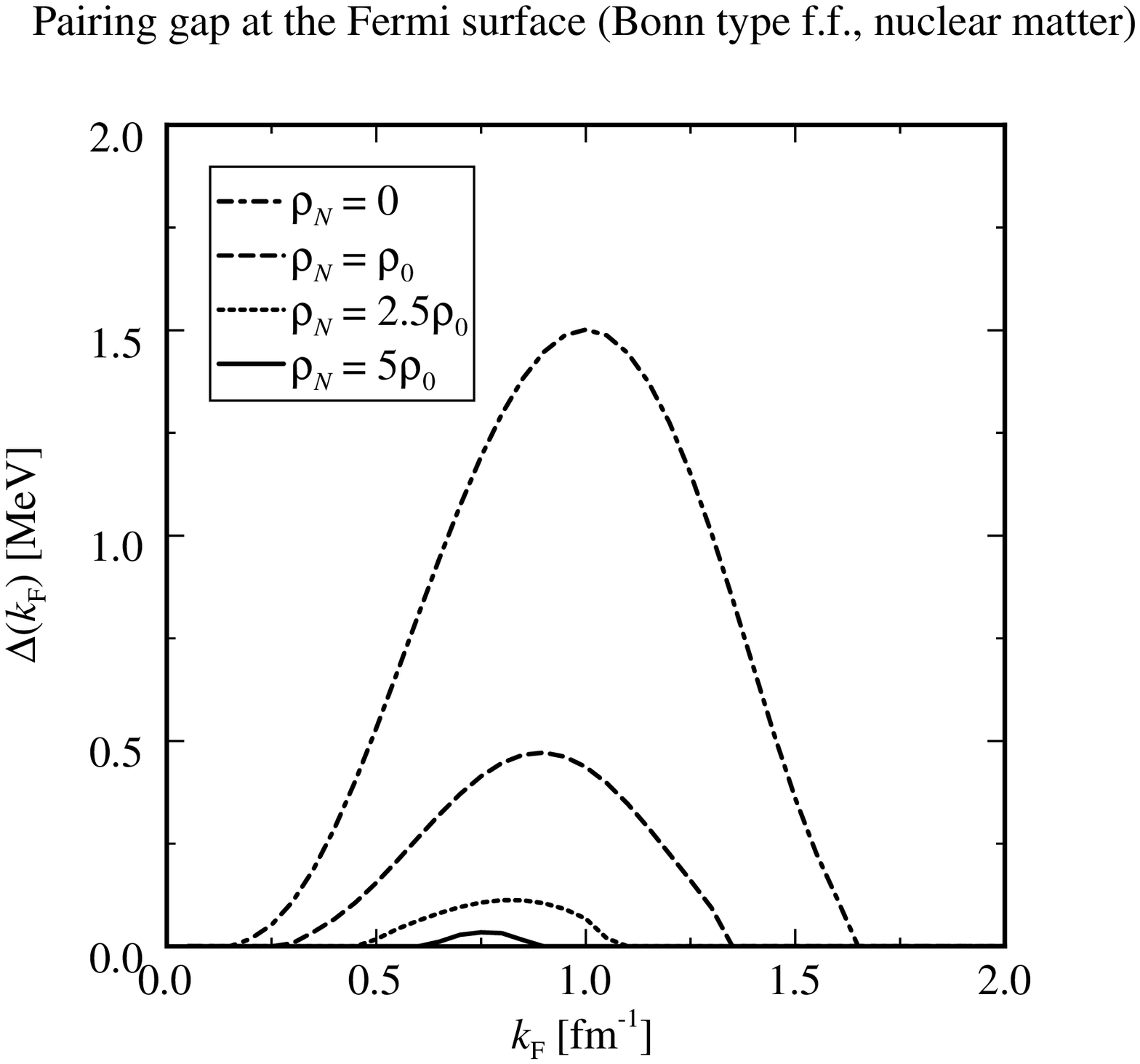} 
  \caption{$\Lambda\Lambda$ pairing gap at the Fermi surface of
  $\Lambda$ hyperons, for nucleon background densities
  $\rho_{N}=0$, $\rho_{0}$, $2.5\rho_{0}$, and $5\rho_{0}$. The
  coupling ratio $\alpha_{\sigma^{\ast}}=0.5$ is used.}
  \label{fig:nuclgap}
\end{figure}

Now we make a comparison between relativistic and nonrelativistic
predictions. 
For the comparison with the nonrelativistic results
of~\citeauthor*{balberg98:_s_lambd}~\cite{balberg98:_s_lambd}, we
calculate the $\Lambda\Lambda$ pairing gap in symmetric nuclear
matter. Figure~\ref{fig:nuclgap} represents our result. Slightly
smaller gap than that obtained from the calculation of pure neutron
matter (Fig.~\ref{fig:neutgap}) reflects the smaller Dirac effective
mass of $\Lambda$ hyperons in symmetric nuclear matter than that in
pure neutron matter.

We would like to note two remarkable differences between their result
and ours. One difference is the dependence of the gap on the
background density. Strikingly, ours is opposite to theirs
(\textit{cf.} Fig.~4 of Ref.~\cite{balberg98:_s_lambd}). This is
brought about \emph{directly} and \emph{indirectly} by decrease of the
Dirac effective mass. We intend by the word `directly' that we can grasp the
decrease of the gap through an expression in the weak-coupling
approximation,
\begin{equation}
  \label{eq:weakcpl}
  \Delta(k_\mathrm{F}) \propto \exp\left[{}-\frac{1}{N(k_\mathrm{F})\,
  |\bar{v}(k_\mathrm{F},k_\mathrm{F})|}\right],
\end{equation}
where $N(k_\mathrm{F}) = E_{k_\mathrm{F}}^{(\Lambda)} k_\mathrm{F} / 2
\pi^{2} \hbar^{2}$ is the density of states at the Fermi
surface. Equation~\eqref{eq:weakcpl} shows that the smaller the Dirac
effective mass becomes, the smaller the density of states does, which
makes the gap smaller. Note that we use the approximation for
rough estimation here and the full integration of the gap
equation~\eqref{eq:gapeq} is done throughout the present study.
Meantime, we intend by the word `indirectly' that the gap decreases as the
density increases due to gradual weakening of the attraction in the
\mbox{p-p} interaction which is shown in Fig.~\ref{fig:mexint}.
The other difference is the region of the Fermi momentum of $\Lambda$
hyperons where the gaps are open. While the regions in their result
are similar in all densities presented, Fig.~\ref{fig:nuclgap} shows
that the regions in our result narrow as the background density
increases.

Finally, the result at $\rho_{N}=\rho_{0}$ may have relevance to the
$\Lambda\Lambda$ pairing correlation around the center of
hypernuclei~\cite{takatsuka01:_hyper_super_neutr_star_cores}: We
obtain the maximal gap $\Delta (k_\mathrm{F}=0.9 \mathrm{~fm}^{-1})
\simeq 0.5$ MeV.

Prior to the present study, \citeauthor{elgaroey96:_super}
studied~\cite{elgaroey96:_super} relativistic effects on the neutron
and proton pairing in neutron star matter and made a comparison with a
nonrelativistic result~\cite{elgaroey96:_model_s_bonn}.
Their result shows a large effect of ``minimal
relativity''~\cite{brown69:_nucleon_nucleon_poten_and_minim_relat} on
the $^{3}P_{2}$ neutron pairing while a small one on the $^{1}S_{0}$
proton pairing. They explained that using DBHF single-particle
energies and factors of the minimal relativity are the causes of
much smaller neutron pairing gap.
As for our model, the factor corresponding to the minimal relativity
is already included in the \mbox{p-p} interaction owing to the
normalization of the Dirac spinor, $u^{\dagger}u=1$.

\subsection{Choice of form factor}
\label{sec:choice-form-factor}

Also noteworthy is a form factor: In this subsection, we investigate a
dependence of the gap on the cutoff mass for each type of the form
factor. We use the purely phenomenological form factor at each
$\Lambda$ hyperon-meson vertex to regulate the high-momentum
components of the \mbox{p-p} interaction as in
Ref.~\cite{matsuzaki01:_phenom}. We have chosen the Bonn-type form
factor, Eq.~\eqref{eq:bonnff}, with the cutoff mass
$\Lambda_\mathrm{c}=7.26$ fm$^{-1}$ thus far in this paper.  In
contrast to the \textit{NN} pairing, there has yet been no proper
guide to determine the cutoff mass in the form factors for the
$\Lambda\Lambda$ pairing. We hence resort to borrow the value from our
previous study of the \textit{NN} pairing~\cite{matsuzaki01:_phenom}.

\begin{figure}[tbp]
  \centering
  \includegraphics[bb=124 388 420
  674,width=7.5cm,keepaspectratio,clip]{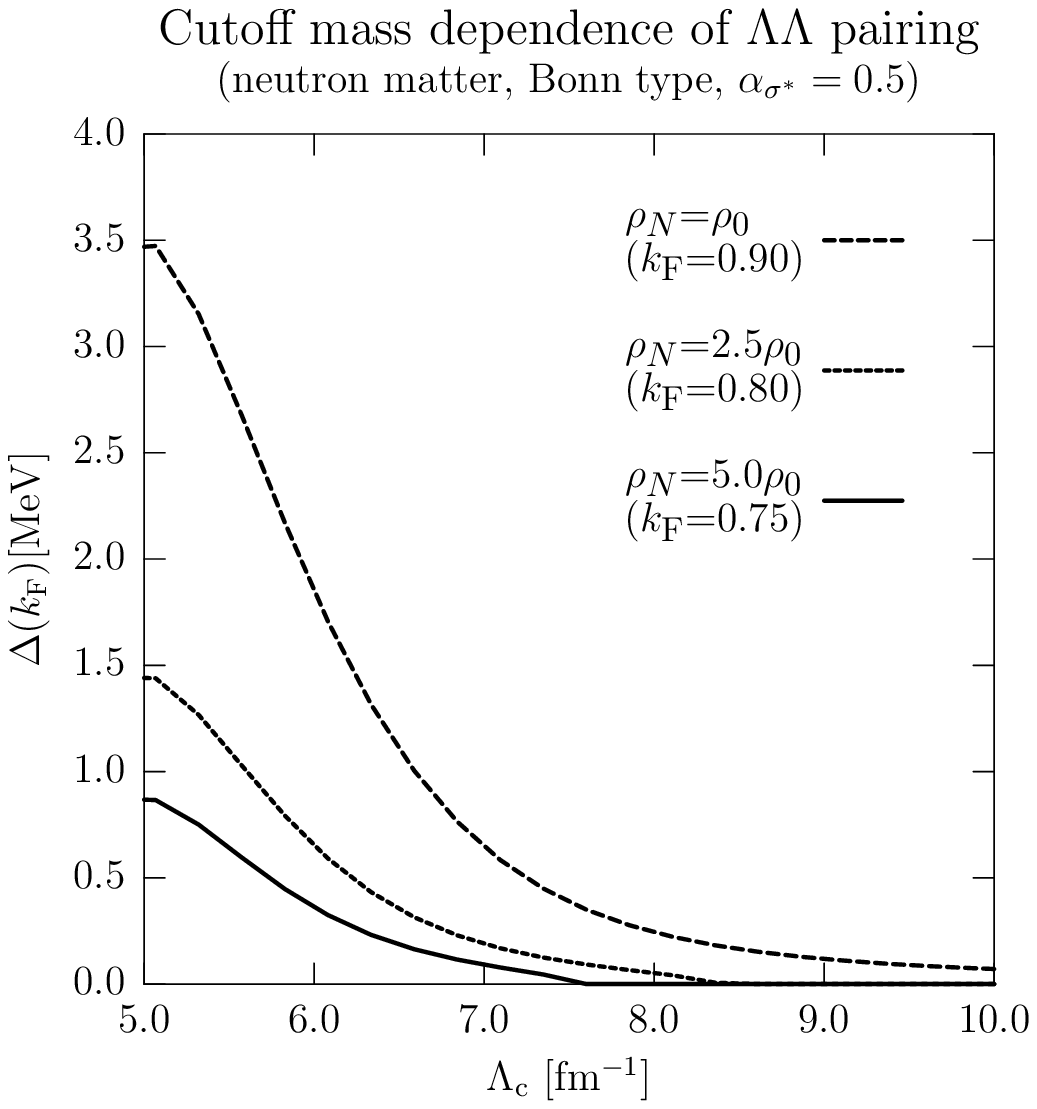}
  \caption{Cutoff mass dependence of the $\Lambda\Lambda$ pairing gap at
    the Fermi surface in pure neutron matter. The coupling ratio
    $\alpha_{\sigma^{\ast}}=0.5$ is used.}
  \label{fig:cutoff-dep}
\end{figure}

However, the type of the form factor and the value of the cutoff mass
significantly affect the magnitude of the pairing gap. We therefore
calculate the dependence of the $\Lambda\Lambda$ pairing gap at the
Fermi surface on the cutoff mass $\Lambda_\mathrm{c}$ in the form
factors of Bonn-type, Eq.~\eqref{eq:bonnff}. The cutoff mass is taken
to be larger than $5$ fm$^{-1}$, which roughly corresponds to mass of
the heaviest meson employed (namely $\phi$), otherwise the interaction
is unphysical. The result is shown in Fig.~\ref{fig:cutoff-dep}, in
which the Fermi momenta of $\Lambda$ hyperons are fixed to
$k_\mathrm{F}=0.90$, $0.80$, and $0.75$ fm$^{-1}$ for background
density of pure neutron matter $\rho_{N}=\rho_{0}$, $2.5\rho_{0}$, and
$5.0\rho_{0}$, respectively.  As expected, varying the cutoff mass
changes the gaps steeply since it changes the balance of the
attraction and the repulsion of the interaction. The peaks around
$\Lambda_\mathrm{c} \sim 5$ fm$^{-1}$ are due to consecutive
suppression of the attraction ($\sigma^{\ast}$ boson) and the
repulsion ($\phi$ meson) by the form factor. Nonetheless, the
importance of this result lies in the fact that the gaps become
smaller in denser background for \emph{any} cutoff mass. Thus the
arbitrariness does not alter our conclusions.

On the other hand, a form factor of monopole type,
\begin{equation}
  \label{eq:monoff}
  f(\mathbf{q}^{2})=
  \frac{\Lambda_\mathrm{c}^{2}}
  {\Lambda_\mathrm{c}^{2}+\mathbf{q}^{2}}\, ,
\end{equation}
with moderate cutoff masses does not give a finite pairing gap in our
model with the HS-m2 set; using other RMF parameter sets may give
finite gaps and their gentle dependence on the cutoff mass is expected
in the manner similar to the \textit{NN}
pairing~\cite{matsuzaki01:_phenom}.

We would like to stress that we do not intend to provide the optimal
parameter sets for the description of $\Lambda\Lambda$ pairing for the
time being; or rather we intend to present its general trend of
density dependence within the present model irrespective of a given
set of parameters. Determining them precisely is inevitably deferred
until the guide can be available.

\subsection{Lambda-Sigma mixing}
\label{sec:lambda-sigma-mixing}

Before concluding the discussions, we present relevant issues for
further study. We have employed pure neutron matter and symmetric
nuclear matter as background in this study. Physics of neutron stars
requires isospin asymmetricity of the background matter which
should be considered in the next study. In connection with this,
coherent and incoherent $\Lambda$-$\Sigma$ couplings should be
mentioned. Reference~\cite{akaishi00:_coher_lambd_sigma_coupl_shell_hyper}
argued that they are important to understand \textit{s}-shell
hypernuclei and, in particular, resolve the longstanding problem of
overbinding  in $\substack{5 \\ \Lambda}$He. Furthermore, the coherent
$\Lambda$-$\Sigma$ coupling predicts the coherent
$\Lambda$-$\Sigma^{0}$ mixing in dense neutron-rich infinite 
matter~\cite{shinmura02:_coher_lambd_sigma}.
As a consequence, the $\Lambda$-$\Sigma^{0}$ mixing shall come into
play in asymmetric nuclear matter, which may change the critical
density of hyperon emergence and eventually scenarios of the
evolution of neutron stars. It is therefore important to introduce it
into the models of dense hadronic matter. Concerning relativistic
models, the introduction into both infinite and finite systems has been
performed using the quantum hadrodynamics along with the concept of
effective field
theory~\cite{mueller99:_effec_lambd_sigma,mueller00:_lambd_sigma};
these works also show the importance of the mixing.
On the other hand, the QCD sum rules predict that relatively weak
mixing would be realized for $\Lambda$ and $\Sigma^{0}$ of the
positive energy state while strong for the
negative~\cite{yagisawa02:_in_sigma_lambd_qcd}. Hence, room for
argument over this issue still remains and the effect of the mixing on
the pairing is unknown so far. In all cases, we have ignored the
mixing since it is beyond our scope of
this study.

\section{Summary}
\label{sec:summ}

We have investigated the $\Lambda\Lambda$ pairing in binary mixed
matter of nucleons and $\Lambda$ hyperons. Our theoretical framework
is the RHB model combined with the RMF interaction both in the
particle-hole and the particle-particle channels; we have used it to
naturally incorporate the medium effect into the latter, as well as
the former, via the Dirac effective mass of $\Lambda$ hyperons in the
$\Lambda$ spinor. Two noteworthy conclusions are thereby drawn.
First, we have found that the value of the $\Lambda\Lambda$ pairing
gap decreases as the background nucleon density increases. This result
is opposite to that reported in Ref.~\cite{balberg98:_s_lambd}. It
should be emphasized again that the origin of the medium effects is
different from each other.
Second, in concert with the effect of increasing the background
density, the weaker the $\Lambda\Lambda$ attraction becomes, the more
the $\Lambda\Lambda$ pairing gap gets suppressed: The present model shows
the possibility that it may disappear in the sequel.  Reluctantly,
some arbitrariness of the form factors still remains since it has yet
been virtually difficult to determine the cutoff mass precisely.  The
magnitudes of the $\Lambda\Lambda$ pairing gap consequently remain
uncertain because they have the strong dependence on the cutoff mass.
We, however, have shown the essential result that the denser
background reduces them gradually to be unchanged for different values
of the cutoff mass.
For transparency of the investigation, we have ignored the chemical
equilibrium. This should be considered in future.
Unfortunately, our knowledge of the hyperon-hyperon interaction is
somewhat limited at the moment.  We notwithstanding expect that
qualitative trends presented in this study will survive in more
refined models, and also in pairing of another hyperonic species.

\newpage
\vspace{\baselineskip}
One of us (T.T.) is grateful to the Japan Society for the
Promotion of Science for research support and the members of the
research group for hadron science at the Japan Atomic Energy Research
Institute (JAERI) for fruitful discussions and their hospitality.

\end{document}